\documentclass{kluwer}    

\usepackage[]{graphicx}
\newdisplay{guess}{Conjecture}

\begin{document}                                                               
\begin{article}
\begin{opening}         
\title{Mid-Infrared observations of GRS\,1915+105 and SS433} 
\author{Ya\"el \surname{Fuchs}$^{(1)}$ \email{yfuchs@cea.fr}}  
\author{I. F\'elix \surname{Mirabel}$^{(1)(2)}$\\}
\author{Richard N. \surname{Ogley}$^{(1)}$}
\runningauthor{Ya\"el Fuchs}
\institute{(1) DSM/DAPNIA/Service d'Astrophysique, CEA/Saclay,\\
Orme des Merisiers B\^at.\,709, 91191 Gif sur Yvette cedex, France\\
(2) IAFE/CONICET, Argentina}
\begin{abstract}
We have observed in the mid-infrared (4-18\,$\mu$m) the counterpart of the 
compact object GRS\,1915+105, and the western jet of SS433. 
The images were carried out with the ISOCAM infrared
camera on board of the Infrared Space Observatory (ISO). 
The mid-infrared photometry of GRS\,1915+105 shows the presence of 
an additional contribution besides the synchrotron emission.
The 15\,$\mu$m images of the large-scale western lobe of the SS433/W50 nebula 
are compared to the radio and X-ray ones. They show infrared synchrotron 
emission on the western edge of SS433/W50 lobe.
\end{abstract}
\end{opening}           
\vspace*{-0.2cm}
\section{GRS\,1915+105}  
We have imaged the close environment of GRS\,1915+105 
with ISOCAM at two epochs, on 1996 April 28 and on 1997 October 24. 
The binary system appears in mid-infrared wavelengths as a faint point source,
globally brighter in our 1997 observations than in the 1996 ones.
No elongated structure was observed with 1.5$''$ pixel field of view, 
the best ISOCAM resolution. 
We obtained the photometry with several large-band filters
between 4 and 18\,$\mu$m, and have compared it to 
simultaneous observations at near-infrared, radio and X-ray wavelengths.

On 1996 April 28 GRS\,1915+105 was in a High/Soft state, with 
nearly constant RXTE/ASM flux ($\sim$ 90\,counts/s), 
no significant BATSE flux,
and $\sim$\,3\,mJy at 15\,GHz (Ryle Telescope) few days before and after. 
On 1997 October 24, 5 days before a major radio and X-ray flare, the compact 
source was in a plateau state with $\sim$ 50\,mJy at 2.25 \& 8\,GHz (GBI)
giving a radio spectral index $\sim -0.08$, 
with low constant RXTE/ASM flux ($\sim$ 20\,counts/s) 
and BASTE photon flux $\sim$ 0.06.

The IR data have been de-reddened according to the law from 
Lutz et al.\,(1996).
We used very simple models combining 
a hot black body emission or a hot synchrotron one to fit the near infrared 
data, with a cold black body emission (for a dust envelope heated by the 
binary) and a synchrotron emission extrapolated from radio wavelengths.
We conclude that synchrotron emission alone is not sufficient to explain the 
mid-infrared flux of GRS\,1915+105, and that an additional contribution, 
which could be dust emission around the binary, is needed.

\section{SS433 / W50}
SS433 is a high mass X-ray binary surrounded by the supernova remnant W50, 
a bright radio nebula with a $\sim$\,1$^{\circ} \times 2^{\circ}$ unusual 
ellipsoidal morphology. 
SS433 produces relativistic jets at subarcsec scales and 
large scale lobes ($\sim$ 90\,pc at a distance of 5\,kpc). 
We have imaged the large scale west radio lobe with the ISOCAM 14-16\,$\mu$m 
filter and a $6''$ pixel field of view resolution in September-October 1997.
\vspace*{-0.3cm}
\begin{figure}[!h]
\centerline{\includegraphics[angle=-90,width=11.8cm]{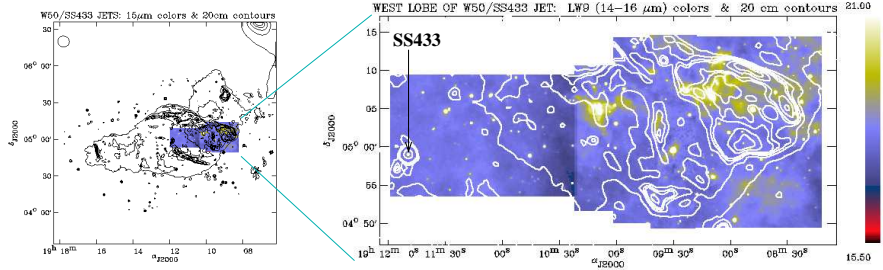}}
\vspace*{-0.2cm}
\caption[]{{\bf Left:} SS433/W50 image at 20\,cm from Dubner et al.\,(1998). 
{\bf Right:} superimposing radio 20\,cm contours with ISOCAM 15\,$\mu$m image 
shows that the IR emission matches the radio one along the western edge of 
the radio nebula. 
{\bf This IR emission is likely to be due to synchrotron radiation.}}
\end{figure}
\vspace*{-0.5cm}
\begin{figure*}[!h]
\centerline{\includegraphics[angle=-90,width=7.7cm]{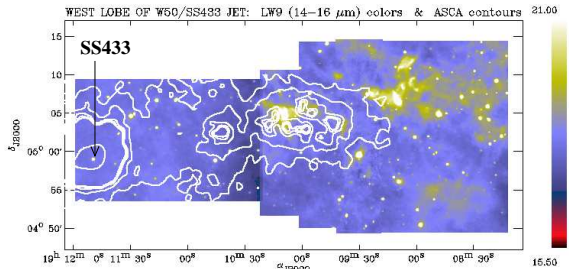}}
\vspace*{-0.2cm}
\caption[!h]{Superimposition of the west lobe X-ray ASCA emission in 
contours with the 15\,$\mu$m image in colors.
It
shows that the IR emission lies along the same axis as the X-ray western lobe.
Thus both emissions are the result of the interaction of the mass outflow with 
the interstellar medium.
Most of the IR emission occurs at the western edge of W50 whereas 
the X-ray lobe fills the space between SS433 and this edge, 
which is an indication that the jet is present between these two components 
without being seen (Mirabel \& Rodr\'\i guez 1999).}
\end{figure*}
\vspace*{-0.5cm}

\end{article}


\begin{thebibliography}{}
\bibitem[]{}Dubner G.M., et al., 1998, {\it ApJ} {\bf 116}, 1842--1855
\bibitem[]{}Lutz D., et al., 1996, {\it A\&A} {\bf 315}, L269--L272
\bibitem[]{}Mahoney W.A., et al., 1997, Proc. of the Fourth Compton Symp. {\it AIP Conf. Proc.} {\bf 410}, pp.-912--916
\bibitem[]{}Margon B., 1984, {\it Annu. Rev. Astron. Astrophys.} {\bf 22}, 507--536
\bibitem[]{}Mirabel I.F. \& Rodr\'\i guez L.F., 1999, {\it Annu. Rev. Astron. Astrophys.} {\bf 37}, 409--443
\end{thebibliography}
\end{document}